\documentclass[a4paper]{PoS}
\usepackage[T1]{fontenc}
\usepackage{mathptmx}
\usepackage{graphicx}	
\usepackage{amsmath}	
\usepackage{amssymb}	
\usepackage{footnote}
\usepackage{subfigure}
\usepackage{subfig}

\title{Constraining GRBs pseudo-redshifts using different empirical correlations}

\ShortTitle{GRBs pseudo-redhisfts}

\author{\speaker{J. Rodrigo Sacahui}\\
        Instituto de Investigaci\'on en Ciencias F\'isicas y Matem\'aticas, USAC, Guatemala.\\
        E-mail: \email{jrsacahui@gmail.com}}

\author{M. Antonio Morales\\
        Instituto de Investigaci\'on en Ciencias F\'isicas y Matem\'aticas, USAC, Guatemala.\\
        E-mail: \email{fantonyovalle@gmail.com }}

\author{M. Magdalena Gonz\'alez\\
        Instituto de Astronom\'ia, Universidad Nacional Aut\'onoma de M\'exico\\
        E-mail: \email{magda@astro.unam.mx}}

\abstract{The determination of distances is highly constrained to a small number of Gamma-Ray Bursts (GRBs) because it requires observations at different wavelengths. Some empirical functions to estimate redshifts have been identified using populations of GRBs with reported redshifts. For example, the Amati correlation relates $E_{peak}$ of the spectrum when modeled with a Band function and the total energy emitted $E_{iso}$ in a time integrated analysis. A multiple-component scenario has been proposed in order to explain GRBs spectra, and in this context when a fine-time spectral analysis is performed a correlation between the non-thermal component's peak energy and the luminosity ($E_{peak,i} - L_i$) appears. This correlation is also used to infer distances to GRBs. In this work we present a sample of bright GRBs and apply these empirical correlations to constrain the pseudo-redshift of the selected burst sample. Our results for GRB080916C, GRB090926A and GRB150214A with reported redshift are totally consistent. Another three bursts with lower luminosities were selected. For these bursts, the pseudo-redshift range obtained with the two correlations are not totally in agreement.}

\FullConference{36th International Cosmic Ray Conference -ICRC2019-\\
		July 24th - August 1st, 2019\\
		Madison, WI, U.S.A.}

\begin{document}

\section{Introduction}\label{sec:introduction}
Gamma Ray Bursts (GRBs) are the most energetic transients in the universe. They are characterized by a prompt release of gamma and X-ray photons with a radiated energy of $\sim 10^{53}$ erg liberated in a few seconds, followed by an long lasting afterglow emission that radiates in different wavelengths. The spectrum of GRBs prompt emission in the keV-MeV energy range is typically described by the empirical Band function \cite{band93} but some bursts present additional spectral components \cite{GO03, GO12, gui11}. The most accepted model describing GRBs emissions is the so called Fireball model \cite{Cavallo78, Paczynski_86, ree94}, which  predicts a compact object as the GRB central engine that can be originated either from the collapse of a massive star \cite{woo93, mac99} or from the merger of two compact objects. This central engine launches high relativistic jets \cite{Paczynski_86}. 

Due to their high luminosity GRBs are visible up to high distances, $z\sim9.4$, becoming good candidates for high redshift studies. Unfortunately despite all the observational an theoretical advances in the field there is not a direct way to infer distances from prompt gamma-ray observations. Different functions (or correlations) have been proposed using prompt, afterglow and prompt-afterglow observations. From prompt emission observations some empirical correlations between GRBs observables have been reported using different spectral components and obtained from both time integrated and fine time analysis, such as the peak energy of the prompt emission spectra, peak luminosity and the isotropic energy \cite{amati06, ghirlanda04, Gui15, gui15a, yonetoku04}.  

In this work we consider two different correlations to infer a redshift range for every GRB of a sample of Fermi bursts using GBM data. First we use one of the most investigated correlation, the so called Amati correlation \cite{amati06}. When fitting the spectrum of the GRB prompt emission with a Band function a correlation is reported between the cosmological rest frame peak energy of the spectrum ($E_{peak}$) and the isotropic equivalent energy $E_{iso}$. Even though this correlation presents a dispersion, it has been used as a pseudo-redshift estimator because of its consistency with several observed redshifts of GRBs from different missions. Amati correlation has been used to constrain cosmological parameters. In the past years several bursts have presented a clear deviation from the Band function in their prompt emission spectra \cite{GO03, GO09, Ack10, gui15a}, in some cases requiring an additional power law component (PL) and in some others a thermal like black body component (BB). 
The second correlation considered in this work is proposed by \cite{Gui15, gui15a, gui16} within a multi-component prompt emission spectral scenario. When fitting the spectra in a fine time spectral analysis of these bursts with three different components: a thermal black body component (BB), a non thermal cutoff power law component (NT) and a Power Law component (PL), a strong correlation in the central engine rest frame appears between the luminosity of the non thermal component of the $i-th$ time interval ($L_i^{NT}$) and the peak energy of the same component in the same time interval. ($E_{peak,i}^{NT}$).

The text is organized as follows. In section {\ref{sec:meth}}, we present the procedure followed for inferring the pseudo-redshifts using the two mentioned correlations. In section {\ref{sec:sample}}, we present the properties of the bursts included in our sample. Finally in section {\ref{sec:results}}, we present the range of values for the distances inferred for the objects in the sample using both correlations and discuss our results.


\section{Empirical correlations}\label{sec:meth}
Most of GRBs prompt emission spectra are best fit with the non-thermal Band function \cite{band93}, two broken power laws smoothly jointed at a break energy. Correlations between some observable quantities have been reported. One of the most studied is the so called Amati correlation. Using a sample of long GRBs with known redshifts a correlation between the rest frame peak energy $E_{peak}$, the energy where the $\nu F\nu$ spectrum presents its maximum, and the isotropic equivalent energy $E_{iso}$ in the energy range between 1-10000 keV has been reported and well studied \cite{amati06}. 

The Amati correlation \cite{amati06} is given as:
\begin{equation}
    E_{p}(\text{keV})= K\times E_{iso}^m(10^{52}\text{erg})
\label{AmatiCorr}
\end{equation}

We use values of $m = 0.57$, $K=80.0$ and a dispersion deviation of the power law of $\sigma =0.18$ as reported in \cite{amati06}. The $E_{peak}$-$E_{iso}$ Amati correlation is cosmological dependent since
\begin{equation}
 E_{iso}(H_o,\Omega_m, \Omega_{\wedge},z )= \frac{4\pi d_L^2(H_o,\Omega_m, \Omega_{\wedge},z)S_{bol}}{1+z}   
\end{equation}
where $d_L(H_o,\Omega_m, \Omega_{\wedge},z)$ is the luminosity distance and $S_{bol}$ is the bolometric fluence of the burst. We use standard cosmological values of Hubble constant $H_o=70$ km s$^{-1}$ Mpc$^{-1}$, matter density parameter $\Omega_m=0.3$ and dark energy parameter $\Omega_{\wedge}=0.7$.

The second correlation used in this work is the one reported by \cite{Gui15, gui15a, gui16}. This one is based on a multi-component spectral scenario: a superposition of a thermal black body component (BB), a non thermal cutoff power law component (NT) and a Power Law component (PL). This correlation is also cosmological dependent and we use the same values for the cosmological parameters as for the Amati correlation. When performing fine time analysis for several Fermi and BATSE bursts a correlation appears between parameters of the non-thermal component, namely the luminosity and the peak energy of the $i-th$ time interval. This correlation is also proposed as a pseudo-redshift estimator and was used to obtain redshift estimates of three BATSE bursts \cite{gui16}. The correlation reported by \cite{Gui15}, as a result of the analysis of several Fermi bursts, is given by:

\begin{equation}
    L_i^{NT}=(9.6\pm 1.1)10^{51}\left( \frac{E_{peak,i}^{rest, NT}}{100 keV} \right)^{1.38\pm0.04} \text{ergs s}^{-1}
\label{SylvRel}
\end{equation}

We perform a fine-time spectral analysis for our selected bursts following the procedure described in  \cite{gui15a} and \cite{gui16}. We assume that the relation \ref{SylvRel} is universal for all Fermi bursts and estimate the corresponding pseudo-redshifts. Roughly, the procedure consists in varying the redshift of the selected burst until the relation of $E_{peak,i}^{NT}$ - $L_i^{NT}$ becomes the relation \ref{SylvRel}. 


\section{Burst Sample}\label{sec:sample}

We performed spectral analysis to a sample of six GRBs detected by Fermi. Three of the selected bursts have reported redshifts. 

GRB 080916C is one of the brightest Fermi burst. It was located at R. A. = 119.85$^o$ and Dec.= -56.63$^o$ with a duration of T$_{90}=$63 s. When fitting this burst with a Band function using three NaI detectors (NaI 0, NaI 3 and NaI 4) along with a BGO detector we obtain spectral parameters in agreement with the ones reported in \cite{Gui15}, peak energy of $E_{peak}= 472 \pm 28$ keV, low energy index $\alpha=-0.9989 \pm 0.0186$ and high energy index of $\beta=-2.309 \pm 0.109$. This burst has a reported redshift of $z=$ 4.15 $\pm$ 0.15 obtained with optical observations by GROND \cite{gre09}.

The very bright burst GRB090926A was observed by Fermi at R. A. = 353.4$^o$ and Dec.= -66.32$^o$ and with a reported redshift of $z=2.1062$ obtained from optical observations by the Very Large Telescope \cite{bis09} of the afterglow emission. We perform the spectral analysis following the procedure of \cite{Gui15} using only GBM data from the two BGO detectors plus three NaI detectors: NaI 3, NaI 6 and NaI 7. When fitting the burst with a Band function we obtained $E_{peak}= 296 \pm 7$ keV, $\alpha=-0.78 \pm 0.02$ and $\beta=-2.43 \pm 0.04$.

GRB150105A was located at R. A. = 124.3$^o$ and Dec= -14.8$^o$ with a duration of T$_{90}=$73 s. In order to use the Amati correlation we performed the time integrated spectral fitting using a Band function. The detectors NaI 8, NaI 11 and BGO1 were used and the resulting parameters are: $E_{peak}= 296 \pm 7$ keV, $\alpha=-0.78 \pm 0.02$ and $\beta=-2.43 \pm 0.04$.

GRB150314A has a reported redshift of $z=$ 1.758 \cite{deUg15}. The integrated spectral analysis was performed using detectors NaI 1, NaI 9 and BGO1. Only a Band function was included. We obtain the following spectral parameters: peak energy of $E_{peak}= 320 \pm 7$ keV, low energy index $\alpha=-0.6087 \pm 0.014$ and high energy index of $\beta=-2.487 \pm 0.0523$.

GRB160113A was located at R. A. = 187.3$^o$ and Dec.= 11.5$^o$ with a duration of T$_{90}=$24 s. It has been associated to possible host candidates within a redshift range between z=0.087 up to z=0.382 \cite{sin16}. We use detectors NaI 8, NaI 11 and BGO1 in the spectral analysis. The time integrated Band function is described by $E_{peak}= 93.3 \pm 1.58$ keV, $\alpha=-0.46 \pm 0.03$ and $\beta=-2.93 \pm 0.09$.

Finally GRB170114A was located at R. A. = 12.08$^o$ and Dec.= -12.55$^o$ with a duration of T$_{90}=$14 s. Detectors NaI 1, NaI 2 and BGO0 were used. The Band parameters of the time integrated spectral fit are $E_{peak}= 217 \pm 15$ keV, $\alpha=-0.77 \pm 0.04$ and $\beta=-2.088 \pm 0.07$.


\section{Results and discussion}\label{sec:results}

\begin{figure}[h]
\centering
\includegraphics[width=0.6\textwidth]{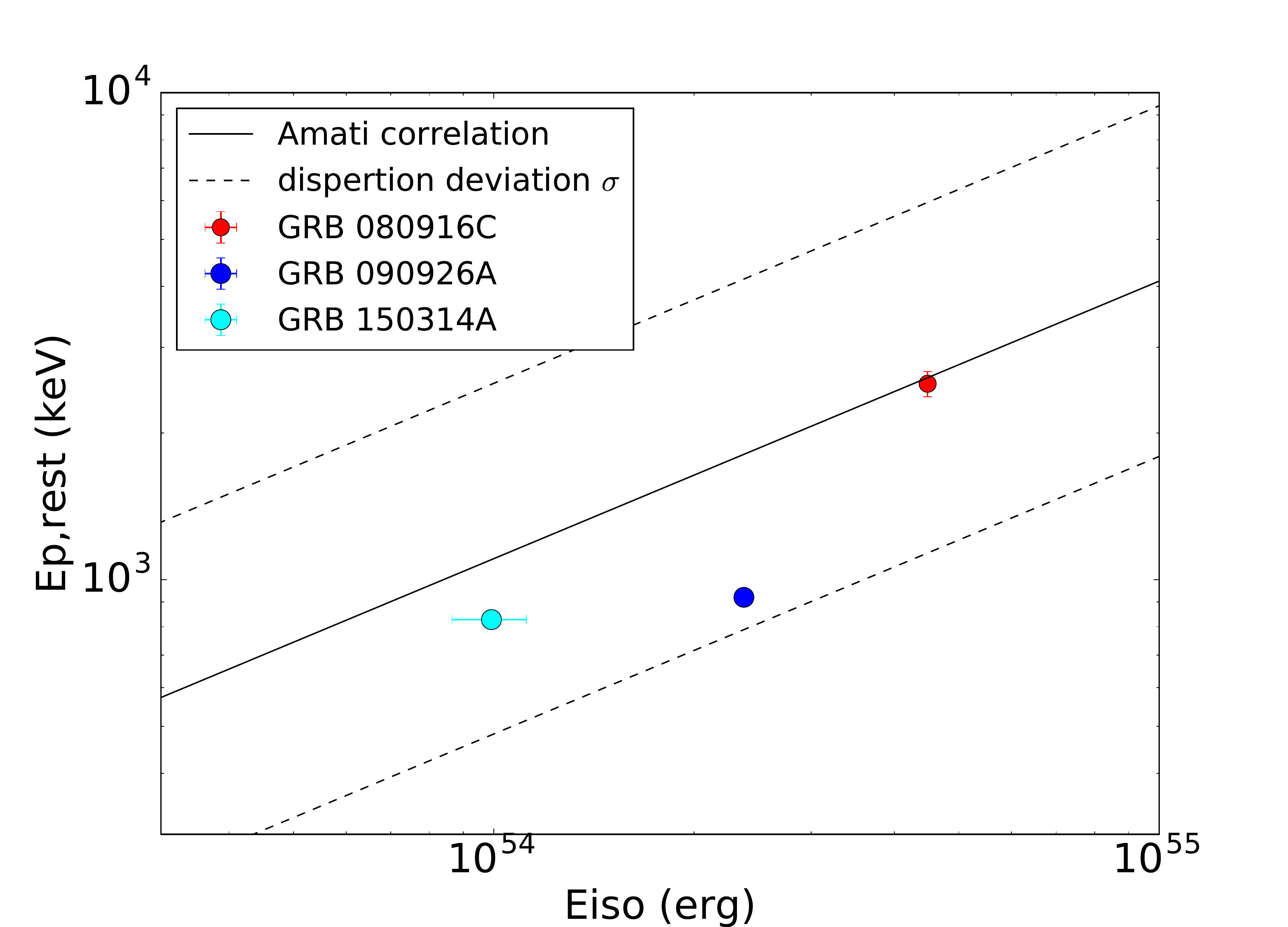}
\caption{GRB080916C, GRB090926A and GRB150314A with reported redshifts follow the Amati correlation.}
\label{AmatiGRBs}
\end{figure}

We follow the two empirical correlations presented in section \ref{sec:meth} in order to obtain pseudo-redshifts of a sample of Fermi GRBs using GBM data. For the three bright GRBs: GRB080916C, GRB090926A and GRB150314A, the results from both correlations are consistent and in agreement with the reported redshifts  (see table \ref{tab:sumary}). In particular, these three bursts are also in agreement with the Amati correlation, see figure \ref{AmatiGRBs},. 

\begin{table}[h]
\centering

\begin{tabular}{cccc} \hline
GRB & Reported  & Amati & redshift inferred  \\ 
 
   & redshift   &  redshift range  &   $E_{peak,i}^{NT}$ - $L_i^{NT} $ correlation \\\hline
GRB080916C & 4.15 $\pm$ 0.15  & > 2  & 3.96 $\pm$ 0.24 \\ GRB090926A & 2.106   & 4 > z > 0.25 & 2.12 $\pm$ 0.16 \\
GRB150314A & 1.758   & z > 0.4 & 1.9 $\pm$ 0.13 \\

\hline

GRB150105A & -- & 0.5 > z > 0.1  & 3.21 $\pm$ 0.25\\
GRB160113A & -- & 1.7 > z > 0.2  & 2.71 $\pm$ 0.2\\
GRB170114A & -- & > 0.7 & 14.31 $\pm$ 0.8\\ \hline

\end{tabular}
\caption{Summary of the pseudo-redshift obtained using the two correlations.}
\label{tab:sumary}
\end{table}

\begin{figure}[h]
\centering
\includegraphics[width=0.6\textwidth]{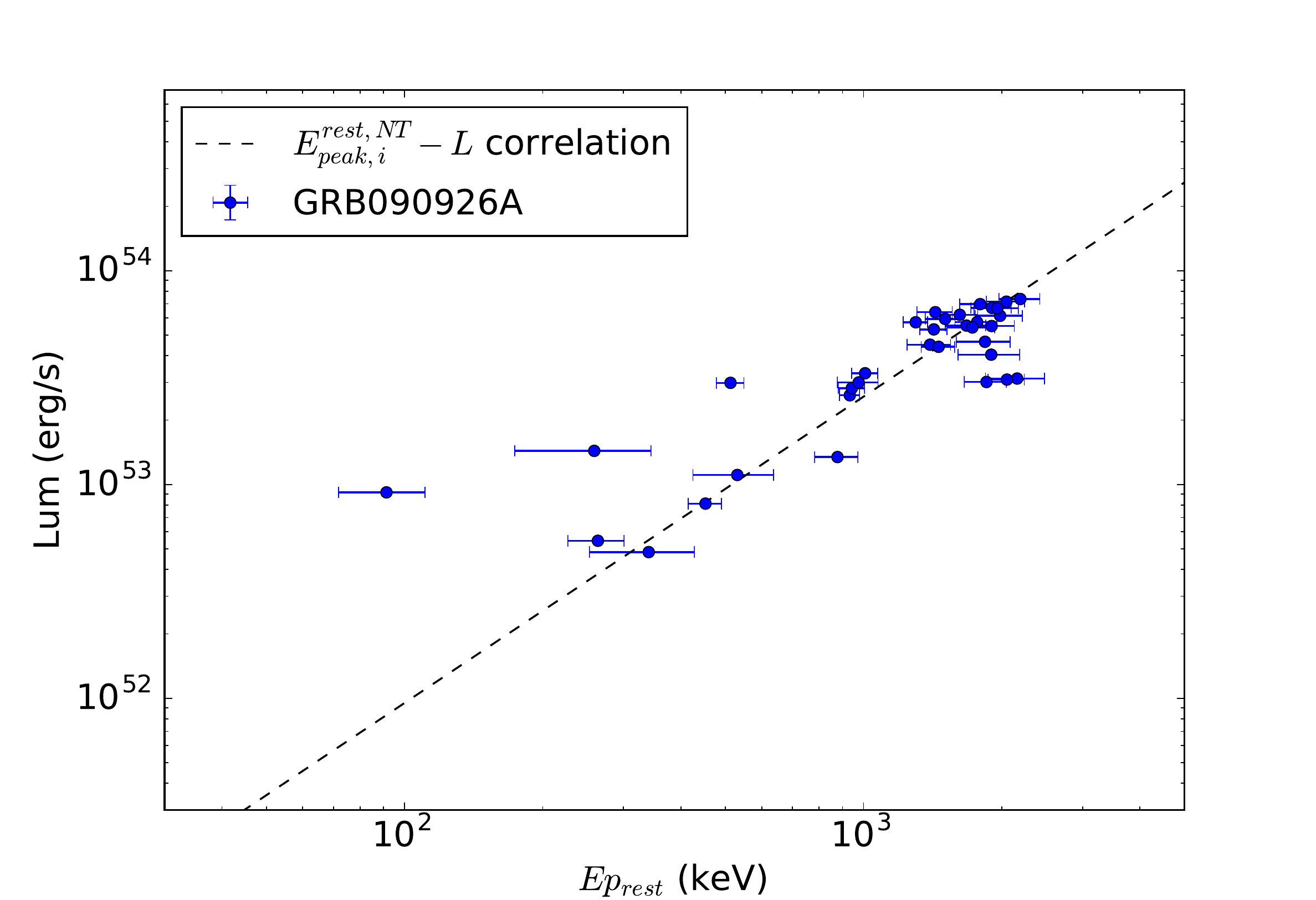}
\caption{$E_{peak,i}^{NT}$ - $L_i^{NT}$ correlation for GRB090926A in a fine time spectral analysis following the procedure presented in \cite{Gui15}.}
\label{dist_grb090926A}
\end{figure}

Burst GRB080916C and GRB090926A present clear deviations from the Band function as reported by \cite{Gui15} and both have been used to test theoretical models. These bursts were used to validate the correlation presented in eq. \ref{SylvRel}. In particular, for GRB 090926A we find the correlation shown in figure \ref{dist_grb090926A} which is in agreement with equation \ref{SylvRel}. 

\begin{figure}[h]
\centering
\subfigure{{\includegraphics[scale=0.33]{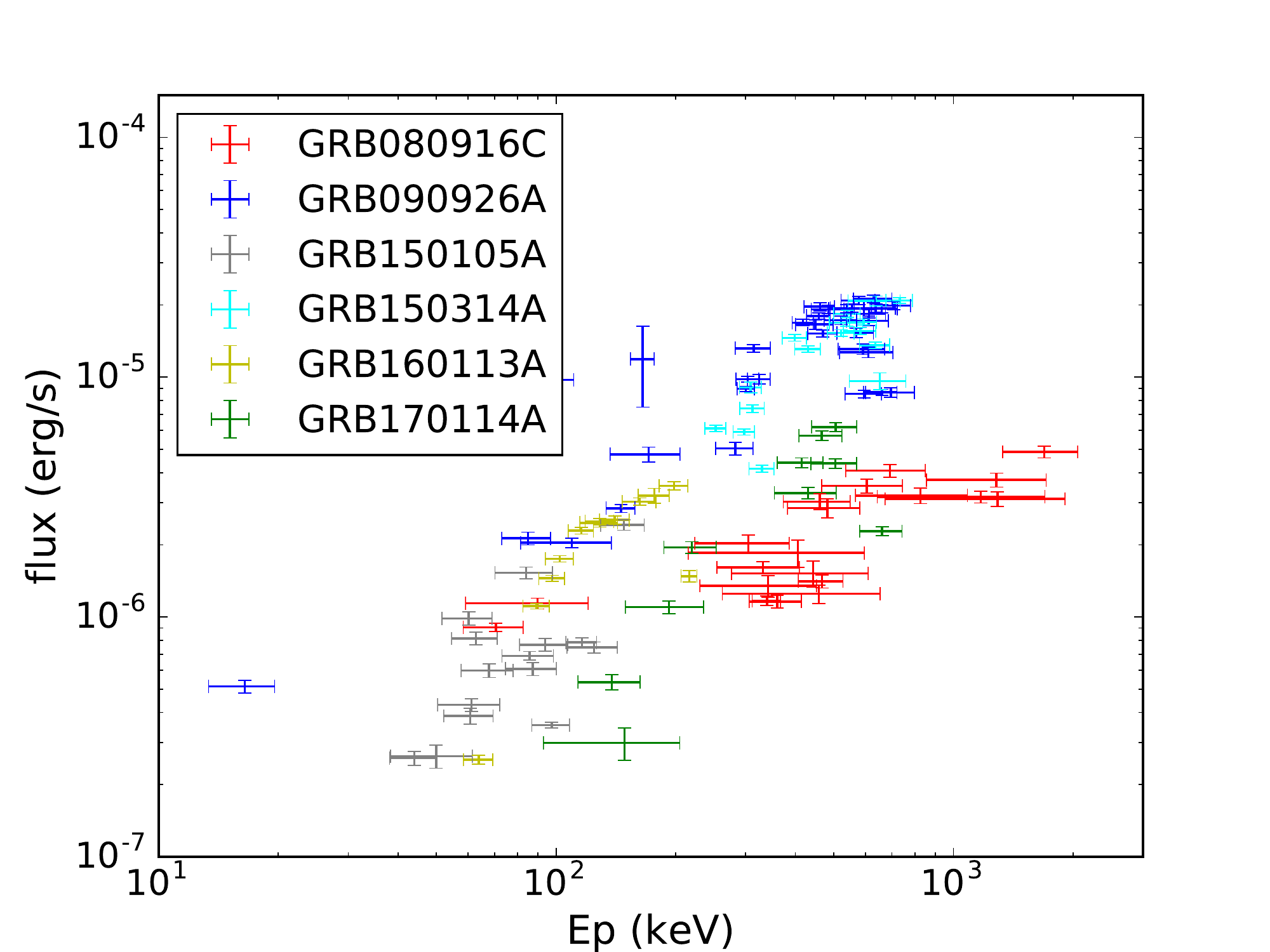} }}%
\hspace{-1.cm}
~
\subfigure{{\includegraphics[scale=0.33]{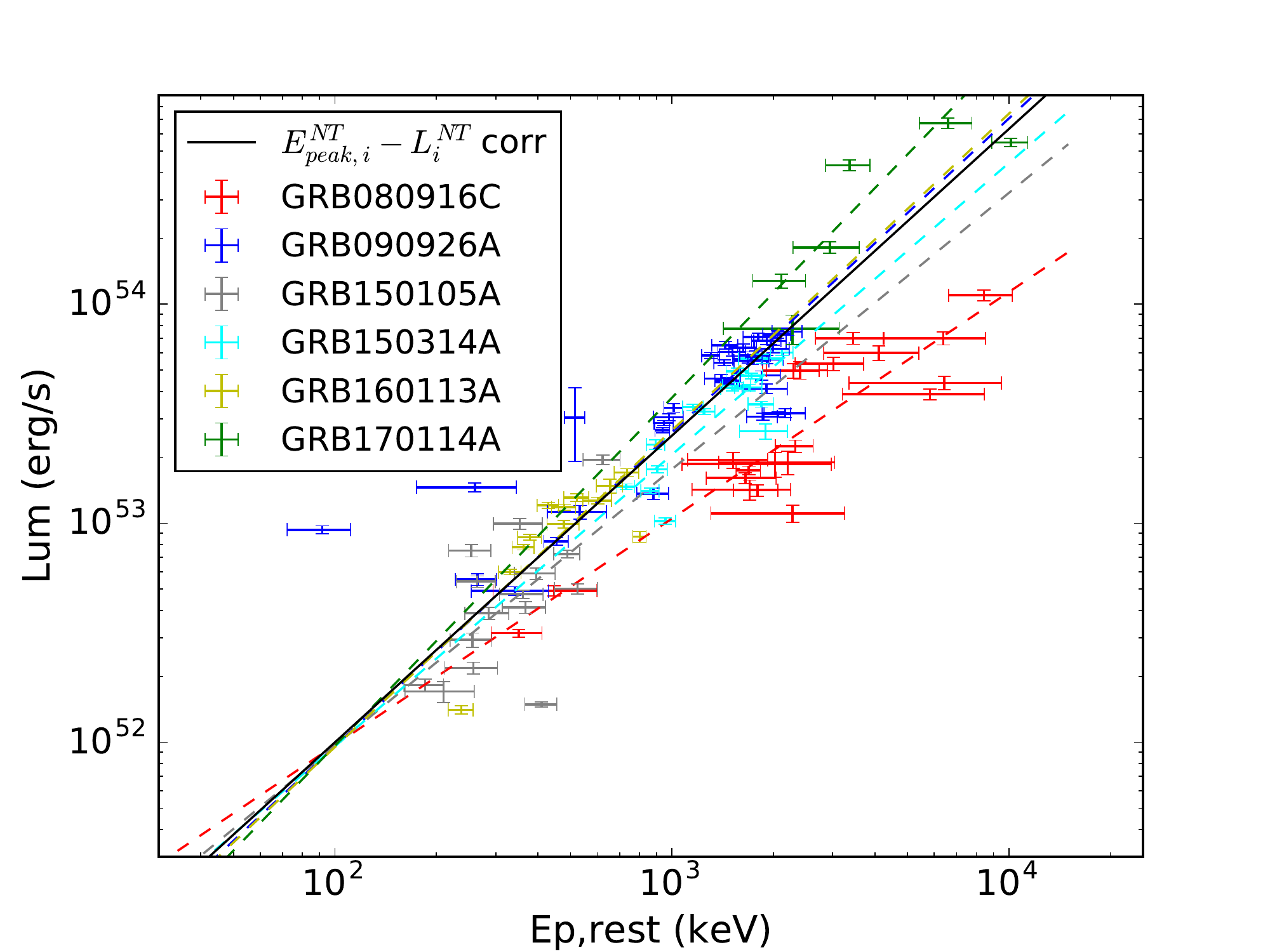} }}%
\caption{In the left panel the plot of E$_{peak}$-Flux of the non thermal component when fitting a three component scenario in a fine time analysis is presented. In the right panel in dotted lines the best E$_{peak,i}^{NT}$-Luminosity at the rest frame that minimize the distance to the relation \ref{SylvRel} used to obtain pseudo-redshifts.}
\label{fig:Ep-Flux}
\end{figure}

To obtain the pseudo-redshift of these bursts using the $E_{peak,i}^{NT}$ - $L_i^{NT}$ correlation we followed the procedure described in section \ref{sec:meth}. First, we obtain the E$_{peak}$-Flux correlation as it is seen in left panel of figure \ref{fig:Ep-Flux}. Then, we obtain the $E_{peak,i}^{NT}$ - $L_i^{NT}$ correlation that is more alike to relation \ref{SylvRel} for each burst, see right panel of figure \ref{fig:Ep-Flux}. 


\begin{figure}[h]
\centering
\subfigure{{\includegraphics[scale=0.35]{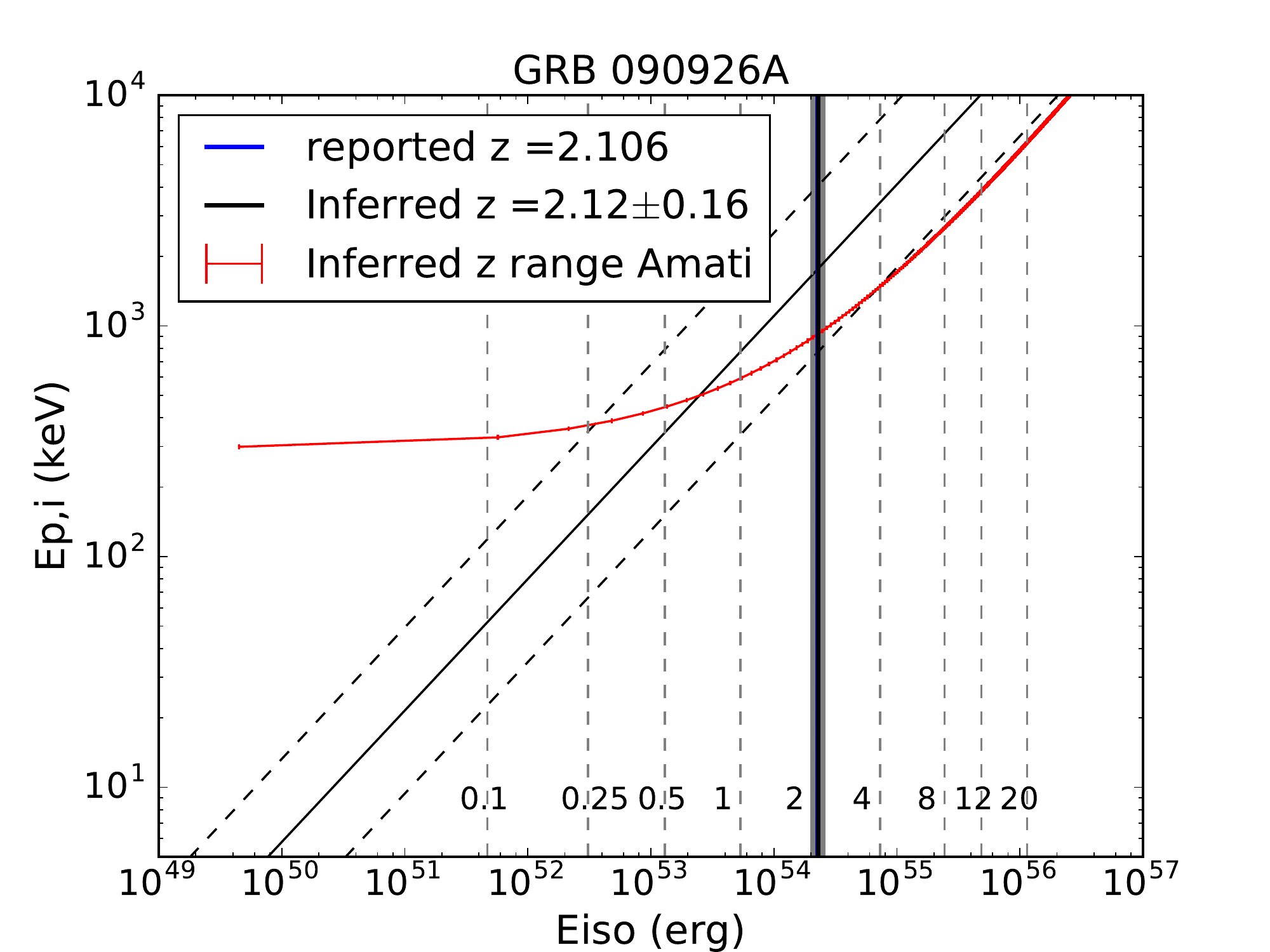} }}%
~
\subfigure{{\includegraphics[scale=0.35]{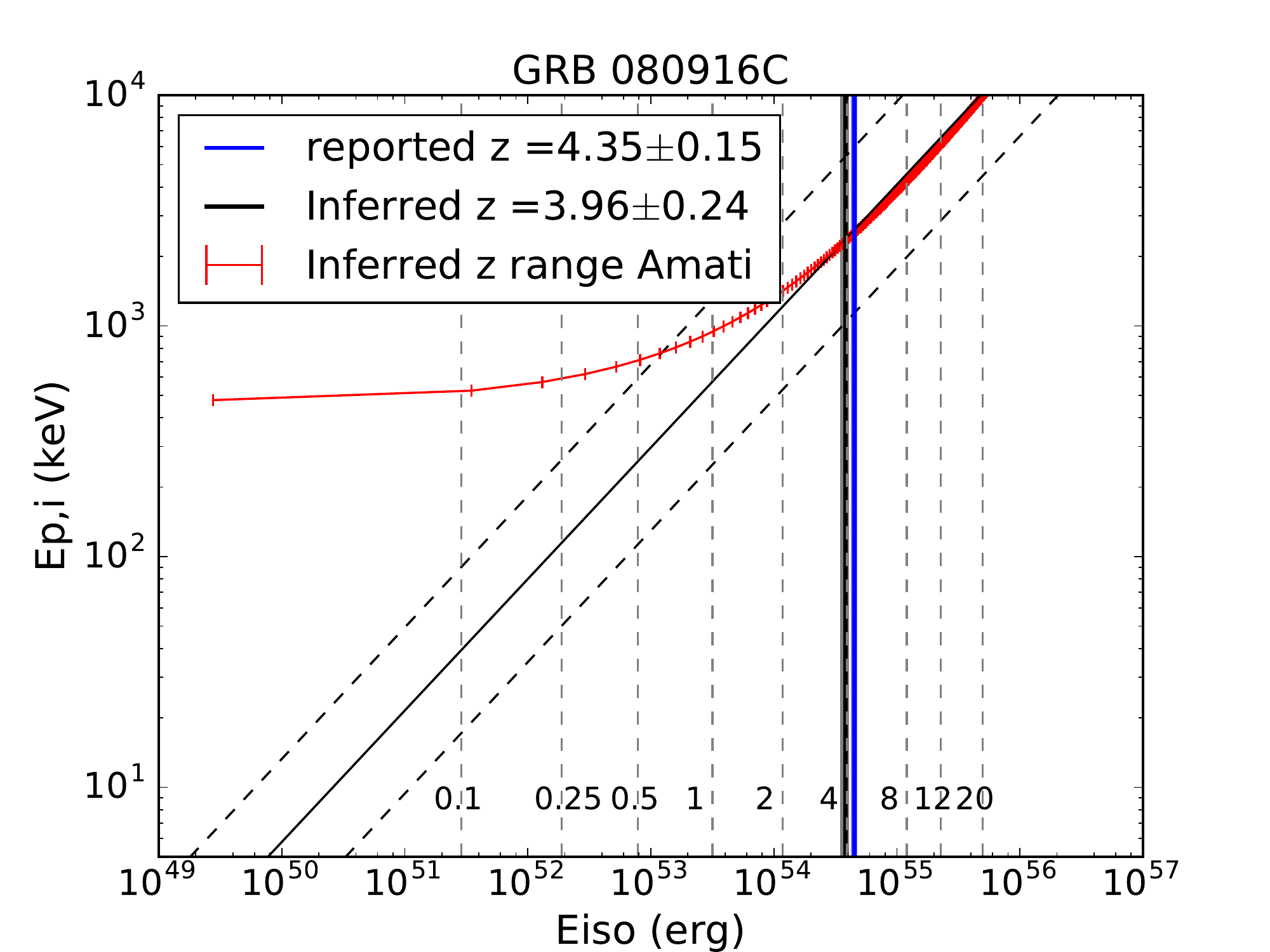} }} \\ \vspace{-0.1cm} %
~
\subfigure{{\includegraphics[scale=0.35]{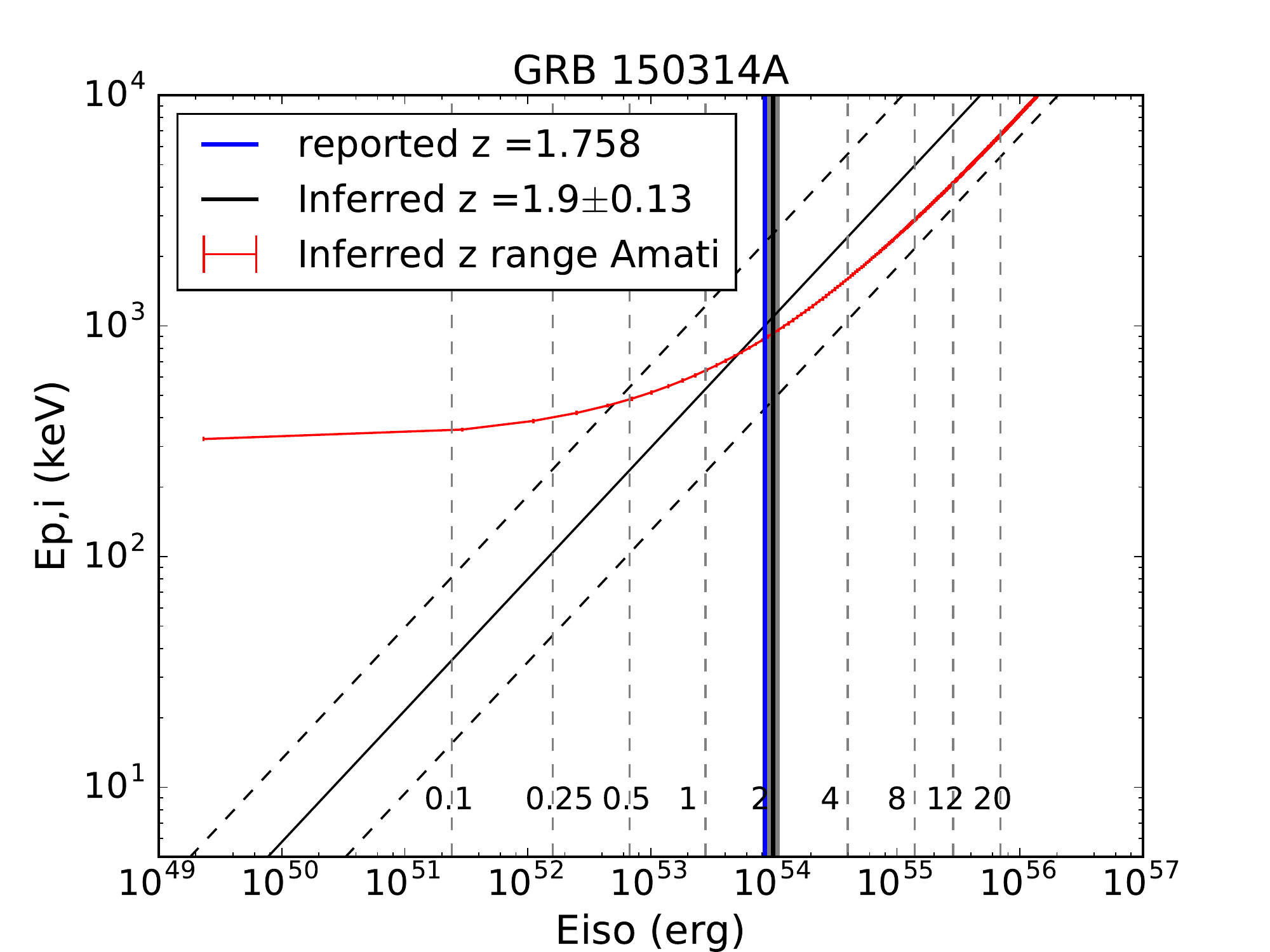} }}%
\caption{The two correlations applied to GRB090926A, GRB080916C and GRB150314A. The pseudo-redshifts obtained with the two correlations used are in agreement with the reported value of redshift obtained from optical observations of their afterglow emission.}
\label{distances}
\end{figure}

For GRB090926A, a value of z = 2.12 $\pm$ 016 was obtained which is really close to the observed value of 2.106. For GRB080916C, the value we obtained for the pseudo-redshift is 3.95 $\pm$ 0.24 which is close to the reported redshift of 4.15 $\pm$ 0.15 and for GRB150314A, a value of 1.9 was obtained, also consistent with the value reported of 1.758. All these values are in the range of pseudo-redshifts obtained for the burst using the Amati correlation as can be seen in figure \ref{distances} and Table \ref{tab:sumary}.

\begin{figure}[h]
\centering
\subfigure{{\includegraphics[scale=0.35]{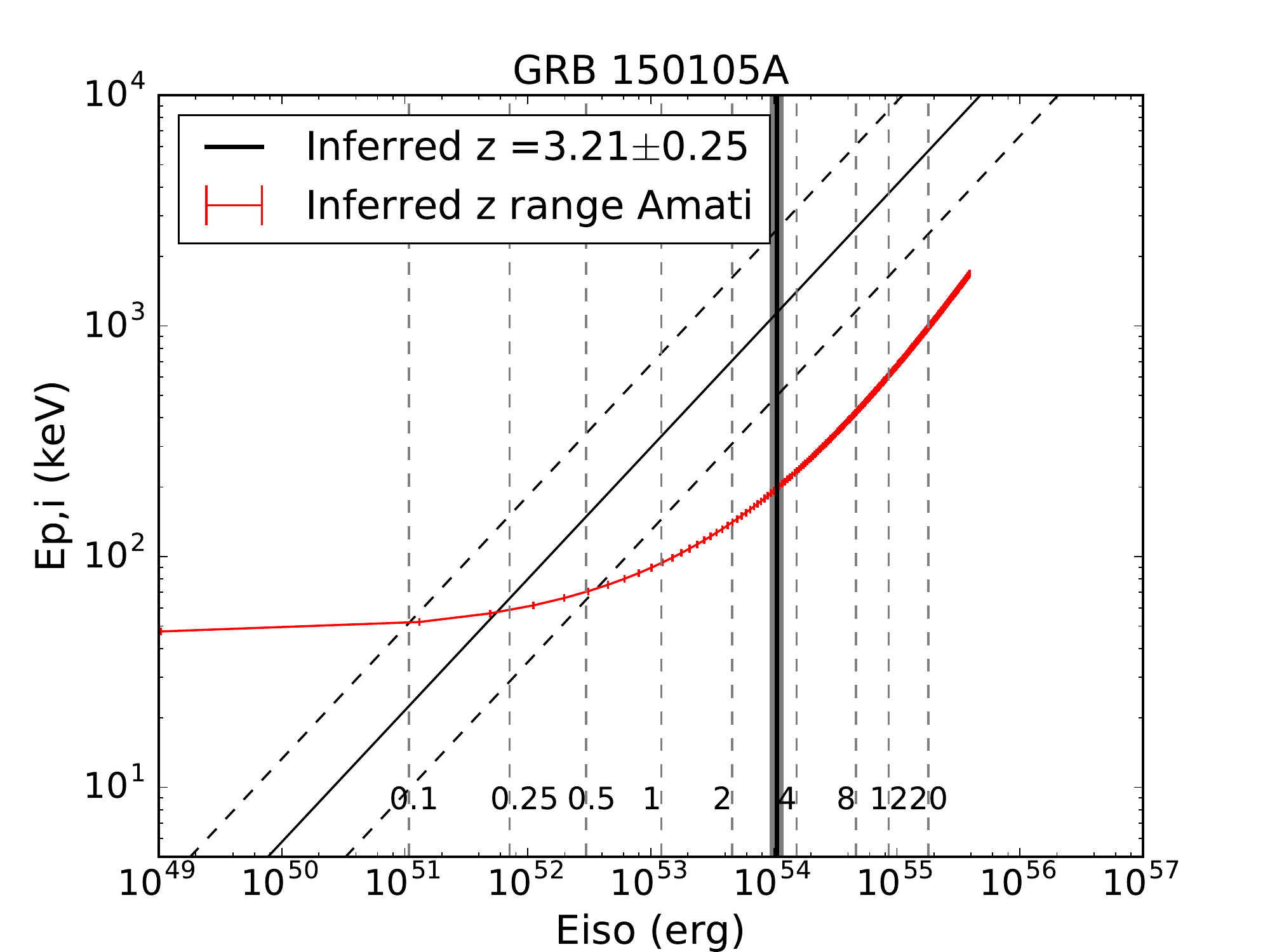} }}%
~
\subfigure{{\includegraphics[scale=0.35]{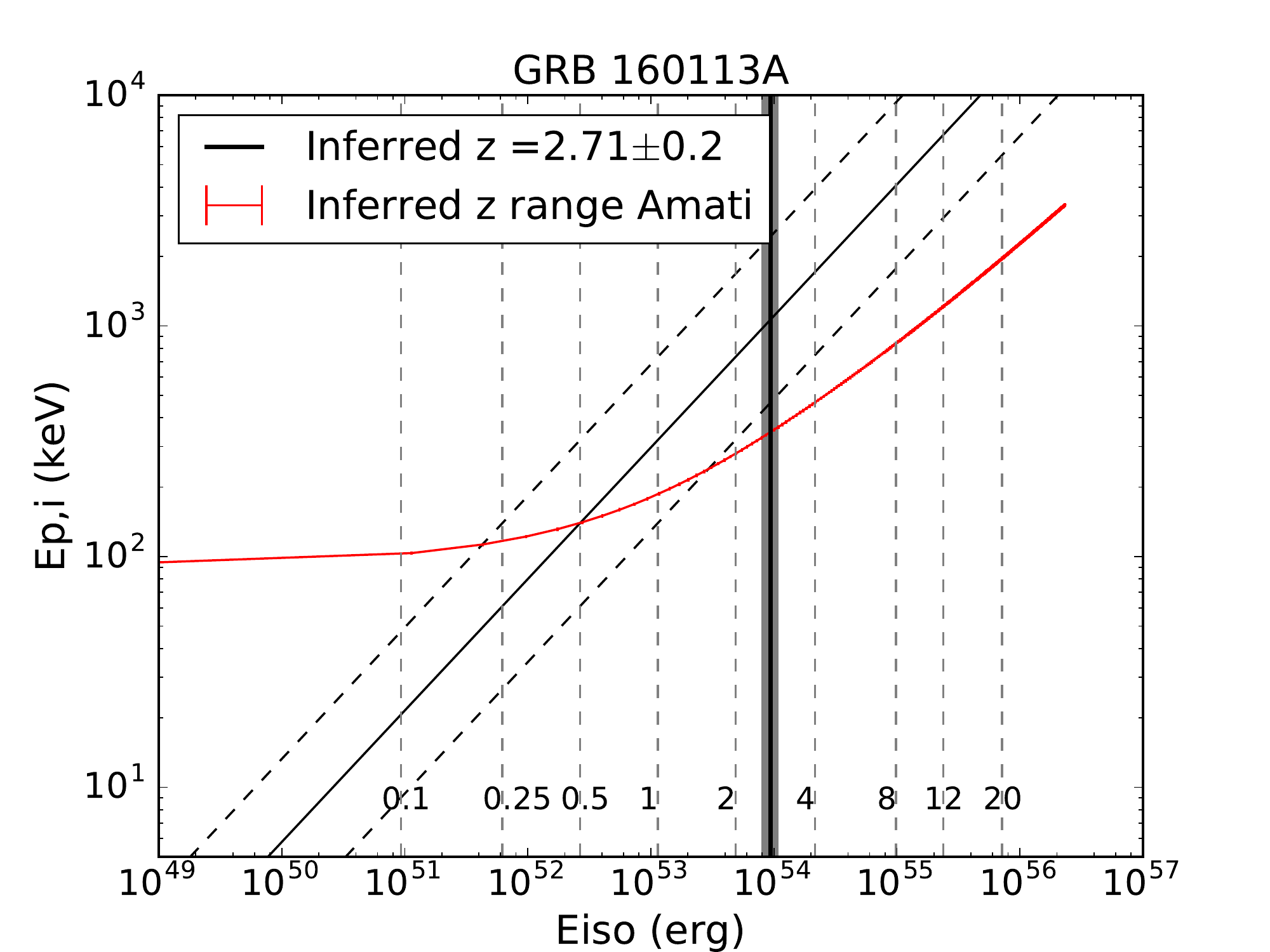} }}\\ \vspace{-0.1cm}%
~
\subfigure{{\includegraphics[scale=0.35]{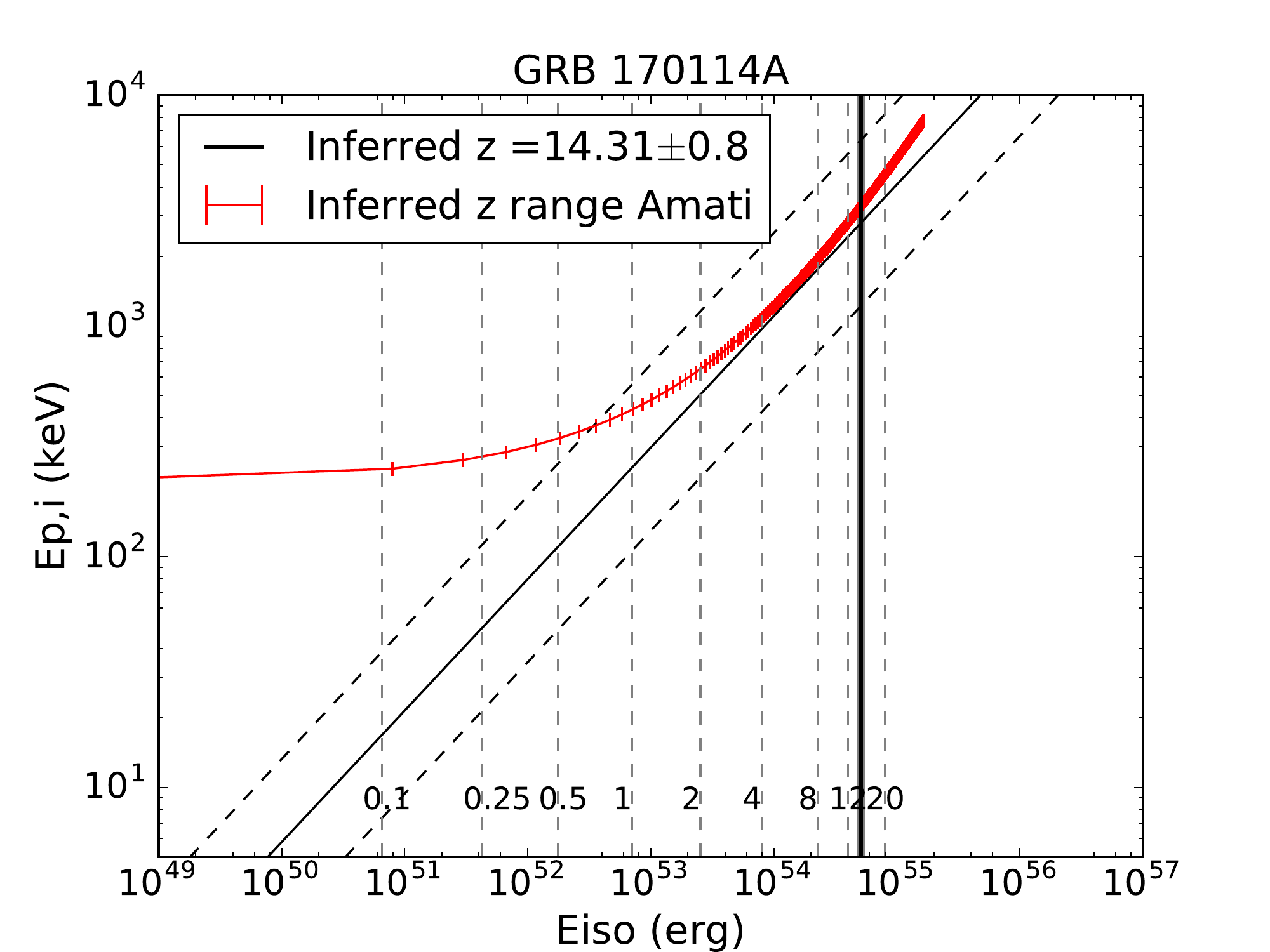} }}%
\caption{The two correlations applied to GRB150105A, GRB160113A and GRB170114A. The pseudo-redshifts obtained are not in total agreement.}
\label{distances_sample}
\end{figure}

For the other three objects the pseudo-redshifts obtained using the two correlations do not seem to be in total agreement as can be seen in figure \ref{distances_sample} and Table \ref{tab:sumary}. This is particularly notorious for GRB150105A for which the Amati approach predicts z between 0.1-0.5 and from relation \ref{SylvRel} a value of 3.21 $\pm$ 0.25 is obtained. They differ by one order of magnitude. 

For GRB160113A there is also disagreement between results from the two correlations. For this burst there are possible host candidates. The redshifts of the hosts obtained from optical observations \cite{sin16} overlap with the results obtained from the Amati correlation. The E$_{peak,i}^{NT}$-Luminosity correlation predicts higher values.

For GRB170114 an extremely high pseudo-redshift is predicted with the E$_{peak,i}^{NT}$-Luminosity correlation.
The values obtained with the Amati correlation are weakly restrictive and include the values obtained by the E$_{peak,i}^{NT}$-Luminosity correlation. This could be the most distant of the three bursts with unknown redshift.

We conclude that the two correlations seems to be in agreement for bright bursts, and that for other less luminous bursts the E$_{peak,i}^{NT}$-Luminosity predicts higher pseudo-redshifts. Several models predict emission in the TeV energy range, high pseudo-redshifts could suggest that a TeV detection is not expected from these bursts. For this sample when considering the E$_{peak,i}^{NT}$-Luminosity correlation all objects were too far to be detected by TeV instruments.

\section*{Acknowledgements}
The authors would like to thank DGAPA-UNAM for financial support by grant AG100317 and DIGED-USAC. 

\bibliographystyle{JHEP.bst}
\bibliography{refgrbs,ReferencesDistances}

\providecommand{\href}[2]{#2}\begingroup\raggedright\begin{thebibliography}{10}

\bibitem{band93}
D.~{Band}, J.~{Matteson}, L.~{Ford}, B.~{Schaefer}, D.~{Palmer}, B.~{Teegarden}
  et~al., \emph{{BATSE observations of gamma-ray burst spectra. I - Spectral
  diversity}}, \href{https://doi.org/10.1086/172995}{\emph{ApJ} {\bfseries 413}
  (1993) 281}.

\bibitem{GO03}
M.~M. {Gonz{\'a}lez}, B.~L. {Dingus}, Y.~{Kaneko}, R.~D. {Preece}, C.~D.
  {Dermer} and M.~S. {Briggs}, \emph{{A {$\gamma$}-ray burst with a high-energy
  spectral component inconsistent with the synchrotron shock model}},
  \href{https://doi.org/10.1038/nature01869}{\emph{Nature} {\bfseries 424}
  (2003) 749}.

\bibitem{GO12}
M.~M. {Gonz{\'a}lez}, J.~R. {Sacahui}, J.~L. {Ramirez}, B.~{Patricelli} and
  Y.~{Kaneko}, \emph{{GRB980923. A Burst with a Short Duration High-energy
  Component}}, \href{https://doi.org/10.1088/0004-637X/755/2/140}{\emph{ApJ}
  {\bfseries 755} (2012) 140}
  [\href{https://arxiv.org/abs/1205.4073}{{\ttfamily 1205.4073}}].

\bibitem{gui11}
S.~{Guiriec}, V.~{Connaughton}, M.~S. {Briggs}, M.~{Burgess}, F.~{Ryde},
  F.~{Daigne} et~al., \emph{{Detection of a Thermal Spectral Component in the
  Prompt Emission of GRB 100724B}},
  \href{https://doi.org/10.1088/2041-8205/727/2/L33}{\emph{ApJ} {\bfseries 727}
  (2011) L33} [\href{https://arxiv.org/abs/1010.4601}{{\ttfamily 1010.4601}}].

\bibitem{Cavallo78}
G.~{Cavallo} and M.~J. {Rees}, \emph{{A qualitative study of cosmic fireballs
  and gamma-ray bursts}}, {\emph{Month notices ..} {\bfseries 183} (1978) 359}.

\bibitem{Paczynski_86}
B.~{Paczynski}, \emph{{Gamma-ray bursters at cosmological distances}},
  \href{https://doi.org/10.1086/184740}{\emph{ApJ} {\bfseries 308} (1986) L43}.

\bibitem{ree94}
M.~J. {Rees} and P.~{Meszaros}, \emph{{Unsteady outflow models for cosmological
  gamma-ray bursts}}, \href{https://doi.org/10.1086/187446}{\emph{ApJ}
  {\bfseries 430} (1994) L93}
  [\href{https://arxiv.org/abs/arXiv:astro-ph/9404038}{{\ttfamily
  arXiv:astro-ph/9404038}}].

\bibitem{woo93}
S.~E. {Woosley}, \emph{{Gamma-ray bursts from stellar mass accretion disks
  around black holes}}, \href{https://doi.org/10.1086/172359}{\emph{ApJ}
  {\bfseries 405} (1993) 273}.

\bibitem{mac99}
A.~I. {MacFadyen} and S.~E. {Woosley}, \emph{{Collapsars: Gamma-Ray Bursts and
  Explosions in ``Failed Supernovae''}},
  \href{https://doi.org/10.1086/307790}{\emph{ApJ} {\bfseries 524} (1999) 262}
  [\href{https://arxiv.org/abs/arXiv:astro-ph/9810274}{{\ttfamily
  arXiv:astro-ph/9810274}}].

\bibitem{amati06}
L.~{Amati}, \emph{{The E$_{p,i}$-E$_{iso}$ correlation in gamma-ray bursts:
  updated observational status, re-analysis and main implications}},
  \href{https://doi.org/10.1111/j.1365-2966.2006.10840.x}{\emph{mnras}
  {\bfseries 372} (2006) 233}
  [\href{https://arxiv.org/abs/astro-ph/0601553}{{\ttfamily
  astro-ph/0601553}}].

\bibitem{ghirlanda04}
G.~{Ghirlanda}, G.~{Ghisellini} and D.~{Lazzati}, \emph{{The
  Collimation-corrected Gamma-Ray Burst Energies Correlate with the Peak Energy
  of Their {\ensuremath{\nu}}F$_{{\ensuremath{\nu}}}$ Spectrum}},
  \href{https://doi.org/10.1086/424913}{\emph{apj} {\bfseries 616} (2004) 331}
  [\href{https://arxiv.org/abs/astro-ph/0405602}{{\ttfamily
  astro-ph/0405602}}].

\bibitem{Gui15}
S.~{Guiriec}, C.~{Kouveliotou}, F.~{Daigne}, B.~{Zhang}, R.~{Hasco{\"e}t},
  R.~S. {Nemmen} et~al., \emph{{Toward a Better Understanding of the GRB
  Phenomenon: a New Model for GRB Prompt Emission and its Effects on the New
  L$_{i}$$^{NT}$- E$_{peak,i}$$^{rest,NT}$ Relation}},
  \href{https://doi.org/10.1088/0004-637X/807/2/148}{\emph{ApJ} {\bfseries 807}
  (2015) 148} [\href{https://arxiv.org/abs/1501.07028}{{\ttfamily
  1501.07028}}].

\bibitem{gui15a}
S.~{Guiriec}, R.~{Mochkovitch}, T.~{Piran}, F.~{Daigne}, C.~{Kouveliotou},
  J.~{Racusin} et~al., \emph{{GRB 131014A: A Laboratory for Studying the
  Thermal-like and Non-thermal Emissions in Gamma-Ray Bursts, and the New
  L$^{nTh}$$_{i}$-E$^{nTh,rest}$$_{peak,i}$ Relation}},
  \href{https://doi.org/10.1088/0004-637X/814/1/10}{\emph{apj} {\bfseries 814}
  (2015) 10} [\href{https://arxiv.org/abs/1507.06976}{{\ttfamily 1507.06976}}].

\bibitem{yonetoku04}
D.~{Yonetoku}, T.~{Murakami}, T.~{Nakamura}, R.~{Yamazaki}, A.~K. {Inoue} and
  K.~{Ioka}, \emph{{Gamma-Ray Burst Formation Rate Inferred from the Spectral
  Peak Energy-Peak Luminosity Relation}},
  \href{https://doi.org/10.1086/421285}{\emph{apj} {\bfseries 609} (2004) 935}
  [\href{https://arxiv.org/abs/astro-ph/0309217}{{\ttfamily
  astro-ph/0309217}}].

\bibitem{GO09}
M.~M. {Gonz{\'a}lez}, M.~{Carrillo-Barrag{\'a}n}, B.~L. {Dingus}, Y.~{Kaneko},
  R.~D. {Preece} and M.~S. {Briggs}, \emph{{Broadband, Time-Dependent,
  Spectroscopy of the Brightest Bursts Observed by BATSE LAD and EGRET TASC}},
  \href{https://doi.org/10.1088/0004-637X/696/2/2155}{\emph{ApJ} {\bfseries
  696} (2009) 2155}.

\bibitem{Ack10}
M.~{Ackermann}, K.~{Asano}, W.~B. {Atwood}, M.~{Axelsson}, L.~{Baldini},
  J.~{Ballet} et~al., \emph{{Fermi Observations of GRB 090510: A Short-Hard
  Gamma-ray Burst with an Additional, Hard Power-law Component from 10 keV TO
  GeV Energies}},
  \href{https://doi.org/10.1088/0004-637X/716/2/1178}{\emph{ApJ} {\bfseries
  716} (2010) 1178} [\href{https://arxiv.org/abs/1005.2141}{{\ttfamily
  1005.2141}}].

\bibitem{gui16}
S.~{Guiriec}, M.~M. {Gonzalez}, J.~R. {Sacahui}, C.~{Kouveliotou}, N.~{Gehrels}
  and J.~{McEnery}, \emph{{CGRO/BATSE Data Support the New Paradigm for GRB
  Prompt Emission and the New L$_{I}$$^{nTh}$-E$_{peak,I}$$^{nTh,rest}$
  Relation}}, \href{https://doi.org/10.3847/0004-637X/819/1/79}{\emph{apj}
  {\bfseries 819} (2016) 79}
  [\href{https://arxiv.org/abs/1507.04081}{{\ttfamily 1507.04081}}].

\bibitem{gre09}
J.~{Greiner}, C.~{Clemens}, T.~{Kr{\"u}hler}, A.~{von Kienlin}, A.~{Rau},
  R.~{Sari} et~al., \emph{{The redshift and afterglow of the extremely
  energetic gamma-ray burst GRB 080916C}},
  \href{https://doi.org/10.1051/0004-6361/200811571}{\emph{aap} {\bfseries 498}
  (2009) 89} [\href{https://arxiv.org/abs/0902.0761}{{\ttfamily 0902.0761}}].

\bibitem{bis09}
E.~{Bissaldi}, \emph{{GRB 090926: Fermi GBM detection.}}, {\emph{GRB
  Coordinates Network} {\bfseries 9933} (2009) 1}.

\bibitem{deUg15}
A.~{de Ugarte Postigo}, J.~P.~U. {Fynbo}, C.~{Thoene}, N.~R. {Tanvir},
  R.~{Sanchez-Ramirez}, J.~{Gorosabel} et~al., \emph{{GRB 150314A: Redshift
  from OSIRIS/GTC.}}, {\emph{GRB Coordinates Network} {\bfseries 17583} (2015)
  1}.

\bibitem{sin16}
L.~P. {Singer}, V.~{Bhalerao}, S.~B. {Cenko} and M.~M. {Kasliwal}, \emph{{GRB
  160113A: iPTF P48 Observations and Optical Transient Candidates.}},
  {\emph{GRB Coordinates Network} {\bfseries 18859} (2016) 1}.

\end{thebibliography}\endgroup



\end{document}